\documentclass {cjpsuppl}         
\usepackage{graphicx}             

\begin{document}
\title{QUANTUM PHASE TRANSITIONS 
       IN $s=\frac{1}{2}$ ISING CHAIN 
       IN A REGULARLY ALTERNATING TRANSVERSE FIELD\,\footnote{Presented 
at 12-th Czech and Slovak Conference on Magnetism, Ko\v{s}ice, 12-15 July 2004}}
\authori{Oleg Derzhko and Taras Krokhmalskii\,\footnote{Corresponding author. 
e-mail address: krokhm@icmp.lviv.ua}}
\addressi{Institute for Condensed Matter Physics NASU,\\ 
          1 Svientsitskii Street, L'viv-11, 79011, Ukraine}
\authorii{}     \addressii{}
\authoriii{}    \addressiii{}
\authoriv{}     \addressiv{}
\authorv{}      \addressv{}
\authorvi{}     \addressvi{}
%
\headauthor{Oleg Derzhko and Taras Krokhmalskii}  
\headtitle{Quantum phase transitions \ldots} 
\lastevenhead{Oleg Derzhko and Taras Krokhmalskii: Quantum phase transitions \ldots} 
%

\refnum{0000 A}
\daterec{XXX}  
\suppl{A}  \year{2004}
\setcounter{page}{1}
\maketitle
\begin{abstract}
We consider the ground-state properties 
of the $s=\frac{1}{2}$ Ising chain 
in a transverse field
which varies regularly along the chain 
having a period of alternation 2.
Such a model, 
similarly to its uniform counterpart,
exhibits quantum phase transitions.
However,
the number and the position of the quantum phase transition points 
depend on the strength of transverse field modulation.
The behaviour in the vicinity of the critical field 
in most cases remains the same as for the uniform chain
(i.e. belongs to the square-lattice Ising model universality class).
However, 
a new critical behaviour may also arise.
We report the results for critical exponents
obtained partially analytically and partially numerically
for very long chains consisting of a few thousand sites.
\end{abstract}

The one-dimensional $s=\frac{1}{2}$ Ising model in a transverse field 
is sometimes called a ``drosophila'' of the quantum phase transition theory 
since it provides the background for the analysis (very often rigorous analytical)
of some important aspects of quantum phase transitions 
which occur in more realistic models of real systems
\cite{[01],[02],[03]}.
In what follows we deal with 
a {\em{regularly alternating}} $s=\frac{1}{2}$ transverse Ising chain
the ground-state and the low-temperature properties of which 
have been discussed recently in Refs. \cite{[04],[05]}. 
To be specific we consider $N\to\infty$ spins $\frac{1}{2}$ 
governed by the Hamiltonian
\bea                        
H=-2\sum_{n} s_n^xs_{n+1}^x 
+\sum_n\left({\mit\Omega} -\left(-1\right)^{n}\Delta{\mit\Omega}\right)s_n^z,
\label{01}
\eea                        
i.e. 
we consider the $s=\frac{1}{2}$ Ising chain 
with the nearest neighbour interaction $J=-2$  
in a regularly alternating transverse field of period 2
the two values of which,
${\mit\Omega}+\Delta{\mit\Omega}$ 
and 
${\mit\Omega}-\Delta{\mit\Omega}$,
are coming into play alternatively.  
The quantum phase transition points for such a chain  
follow from the condition \cite{[06]}
$\left({\mit\Omega}^\star+\Delta{\mit\Omega}\right)
\left({\mit\Omega}^\star-\Delta{\mit\Omega}\right)=\pm 1$,
which immediately yields 
${\mit\Omega}^\star=\pm\sqrt{1+\Delta{\mit\Omega}^2}$
if $\Delta{\mit\Omega}<1$,
${\mit\Omega}^\star=\left\{-\sqrt{2},\;0,\;\sqrt{2}\right\}$
if $\Delta{\mit\Omega}=1$
and
${\mit\Omega}^\star=\pm\sqrt{\Delta{\mit\Omega}^2\pm 1}$
if $\Delta{\mit\Omega}>1$.
Hereafter we focus on the behaviour 
of different ground-state quantities 
of the spin chain (\ref{01}) with $\Delta{\mit\Omega}=1$ 
in the vicinity of critical points 
${\mit\Omega}^\star=\sqrt{2}$
and
${\mit\Omega}^\star=0$
since for other values of $\Delta{\mit\Omega}$
(or for $\Delta{\mit\Omega}=1$ and ${\mit\Omega}^\star=-\sqrt{2}$)
the critical behaviour is the same 
as the one in the vicinity of ${\mit\Omega}^\star=\sqrt{2}$.

A critical behaviour is characterised by a set of exponents 
which determine peculiarities of different ground-state quantities 
in the vicinity of the critical field.
Thus,
the order parameter 
(i.e. the longitudinal magnetization 
$m^x=\frac{1}{N}\sum_n\langle s_n^x\rangle$)
decays as 
$m^x\sim\left({\mit\Omega}^\star-{\mit\Omega}\right)^{\beta}$,
the equal-time two-point correlation function of $x$ spin component 
behaves as 
$\langle s_n^xs^x_{n+2r}\rangle
-
\langle s_n^x\rangle\langle s^x_{n+2r}\rangle
\stackrel{2r\to\infty}{\sim}
\frac{1}{\left(2r\right)^\eta}\exp\left(-\frac{2r}{\xi}\right)$
with the correlation length 
$\xi\sim\left\vert{\mit\Omega}-{\mit\Omega}^\star\right\vert^{-\nu}$.
The transverse static susceptibility diverges as 
$\chi^z\sim\left\vert{\mit\Omega}-{\mit\Omega}^\star\right\vert^{-\alpha}$.
Finally, the energy gap vanishes as
$\Delta\sim\xi^{-z}
\sim\left\vert{\mit\Omega}-{\mit\Omega}^\star\right\vert^{\nu z}$,
where $z$ is the dynamic critical exponent.
The quantum phase transition 
in $d$ space dimensions 
(in our case $d=1$)
corresponds
to the finite-temperature (``classical'') phase transition 
in $d+z$ space dimensions.
Moreover,
from the classical phase transition theory 
we know
a number of relations 
which connect the critical exponents:
$2-\alpha=2\beta+\gamma$,
$2-\alpha=\beta\left(\delta+1\right)$,
$2-\alpha=d\nu$,
$\gamma=\left(2-\eta\right)\nu$.

In the uniform case ($\Delta{\mit\Omega}=0$) 
the critical exponents are well known:
$\beta=\frac{1}{8}$,
$\nu=1$,
$\eta=\frac{1}{4}$,
$\alpha=0$,
$z=1$.
This implies that the quantum phase transition belongs 
to the square-lattice Ising model universality class.
In the considered case of regularly alternating chain (\ref{01})
with $\Delta{\mit\Omega}=1$
we shall see 
that the critical behaviour in the vicinity of ${\mit\Omega}^\star=\sqrt{2}$
remains in the same universality class.
However, 
the critical behaviour in the vicinity of ${\mit\Omega}^\star=0$
is characterised by a different set of exponents.

We start with the exponents $\nu z$ and $\alpha$ 
which can be obtained analytically 
using the Jordan-Wigner fermionization and continued fractions.
From Refs. \cite{[04],[05]} we know the elementary excitation energy distribution
which permits us to get for the energy gap
in the vicinity of the critical points,
$\Delta\sim \vert{\mit\Omega}-\sqrt{2}\vert$
and 
$\Delta\sim\vert{\mit\Omega}\vert^2$.
This implies $\nu z=1$ and $\nu z=2$ 
for the critical points ${\mit\Omega}^\star=\sqrt{2}$ and ${\mit\Omega}^\star=0$,
respectively.
Moreover, 
from Refs. \cite{[04],[05]} we know 
that the ground-state energy contains a nonanalytical contribution, 
$e_0\sim \left({\mit\Omega}-\sqrt{2}\right)^2\ln\vert{\mit\Omega}-\sqrt{2}\vert$
and 
$e_0\sim {\mit\Omega}^4\ln\vert{\mit\Omega}\vert$, 
in the vicinity of ${\mit\Omega}=\sqrt{2}$ and ${\mit\Omega}=0$,
respectively.
After differentiating we conclude  
that
$\chi^z=\frac{\partial^2e_0}{\partial{\mit\Omega}^2}$ 
exhibits a logarithmic divergency at ${\mit\Omega}^\star=\sqrt{2}$
and remains finite at  ${\mit\Omega}^\star=0$ 
where, 
nevertheless, 
$\frac{\partial^2\chi^z}{\partial{\mit\Omega}^2}$ 
exhibits a logarithmic divergency.
Therefore,
$\alpha=0$ at ${\mit\Omega}^\star=\sqrt{2}$
and $\alpha=-2$ at ${\mit\Omega}^\star=0$.

We go on to the exponents $\beta$, $\nu$ and $\eta$
which can be calculated numerically \cite{[07]}.
Knowing the spin correlation function
$\langle s_n^x s_{n+2r}^x\rangle$
we can obtain the on-site magnetizations
$m_n^x=\sqrt{\lim_{r\to\infty}\langle s_n^x s_{n+2r}^x\rangle}$,
$n=1,2$
and 
$m^x=\frac{1}{2}\left(m_1^x+m_2^x\right)$.
Assuming for 
$\langle s_n^x s_{n+2r}^x\rangle -\left(m^x_n\right)^2$
the long-distance behaviour
$\sim\left(2r\right)^{-\eta}\exp\left(-\frac{2r}{\xi}\right)$
as $r\to\infty$
we can find the correlation length $\xi$ and the exponent $\eta$.
To determine $m^x$ (Fig. \ref{fig01}) 
we take $N=2000,\;4000,\;5400$,
$n=\frac{N}{4}$,
$2r=\frac{N}{2}$.
%
%
\bfg[h,t]                  
\vspace{0mm}
\bc                        
\includegraphics[width=120mm]{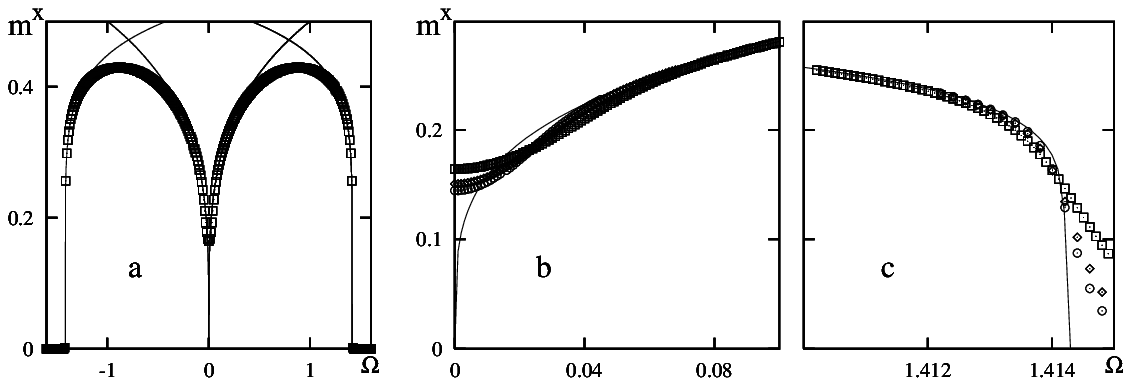}               
\ec                        
\vspace{-3mm}
\caption{The order parameter behaviour in the vicinity of critical points
($N=2000$: squares,
$N=4000$: diamonds,
$N=5400$: circles).
The solid lines in panels a, b and c 
represent the dependences 
$m^x =0.5\left(\sqrt{2}\pm{\mit\Omega}\right)^{\frac{1}{8}}$
and
$m^x=0.5\vert{\mit\Omega}\vert^{\frac{1}{4}}$ (a),
$m^x=0.5{\mit\Omega}^{\frac{1}{4}}$ (b)
and 
$m^x=0.5\left(\sqrt{2}-{\mit\Omega}\right)^{\frac{1}{8}}$ (c).
\label{fig01}}
\efg                       
To determine $\xi$ and $\eta$ (Figs. \ref{fig02}, \ref{fig03}) 
we take 
$N=2000$ and $n=920$,
$N=4000$ and $n=1920$,
$N=5400$ and $n=2620$, 
and $2r=162,170,178,186,194$.
%
%
\bfg[h,t]                  
\vspace{0mm}
\bc                        
\includegraphics[width=120mm]{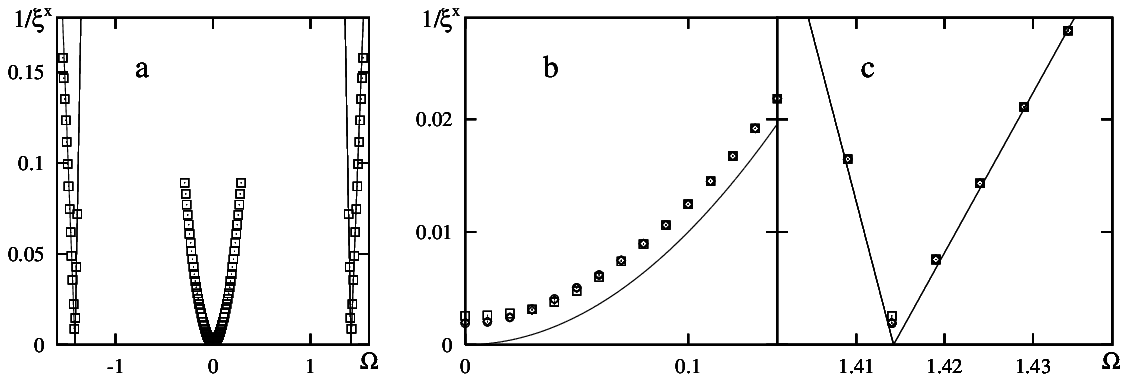}              
\ec                        
\vspace{-3mm}
\caption{The inverse correlation length behaviour in the vicinity of critical points
($N=2000$: squares, $N=4000$: diamonds, $N=5400$: circles;
diamonds coincide with circles in the scale of panels b and c).
The solid lines  
represent the dependences 
$\frac{1}{\xi}=3\vert\sqrt{2}-{\mit\Omega}\vert$ 
if ${\mit\Omega}\le\sqrt{2}$ (a and c),
$\frac{1}{\xi}=\sqrt{2}\vert\sqrt{2}-{\mit\Omega}\vert$ 
if ${\mit\Omega}\ge\sqrt{2}$ (a and c)
and
$\frac{1}{\xi}={\mit\Omega}^2$ (a and b).
\label{fig02}}
\efg                       
%
%
\bfg[h,t]                  
\vspace{0mm}
\bc                        
\includegraphics[width=120mm]{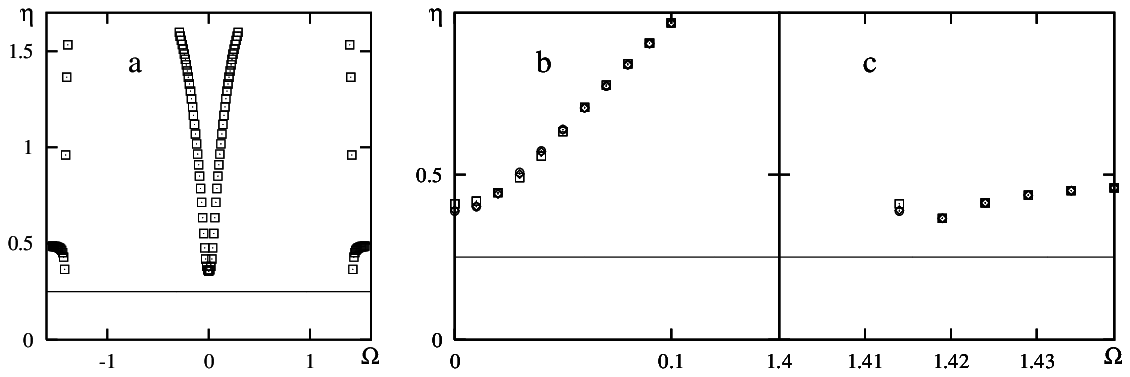}              
\ec                        
\vspace{-3mm}
\caption{The power-law exponent $\eta$ in the vicinity of critical points
($N=2000$: squares, $N=4000$: diamonds, $N=5400$: circles
(they cover diamonds in the scale of panels b and c)).
$\eta$ tends to 0.25 at the critical points very slowly with increasing $N$.
\label{fig03}}
\efg                       

Let us comment on the the obtained results.
From the data reported in Figs. \ref{fig01}, \ref{fig02}, \ref{fig03}
we may conclude 
that the critical behaviour in the vicinity of ${\mit\Omega}^\star=\sqrt{2}$ 
is characterised by the set 
$\beta=0.125$, $\nu=1$ and $\eta=0.25$.
Contrary,
in the vicinity of ${\mit\Omega}^\star=0$
the corresponding exponents are 
$\beta=0.25$, $\nu=2$ and $\eta=0.25$.
Obviously, we expect such $N\to\infty$ values of $\beta$, $\nu$ and $\eta$ 
from the finite-$N$ data 
following the tendencies in their change as $N$ increases.
Combining these findings with the analytical results 
for $\nu z$ and $\alpha$
we get two sets of critical exponents.
Namely,
$\beta=0.125$,
$\nu=1$,
$\eta=0.25$,
$\alpha=0$,
$\nu z=1$
in the vicinity of ${\mit\Omega}^\star=\sqrt{2}$
and
$\beta=0.25$,
$\nu=2$,
$\eta=0.25$,
$\alpha=-2$,
$\nu z=2$
in the vicinity of ${\mit\Omega}^\star=0$.
The former set corresponds 
to the square-lattice Ising model universality class 
whereas the latter one corresponds to a new universality class.

To summarize,
we have studied the quantum phase transitions 
in the $s=\frac{1}{2}$ Ising chain 
in a regularly alternating transverse field of period 2,
${\mit\Omega}\pm\Delta{\mit\Omega}$.
We have found 
that for a small strength of regular nonuniformity 
(controlled by $\Delta{\mit\Omega}$)
the ground-state phase transition between the Ising phase 
(for small ${\mit\Omega}$)
and the paramagnetic phase 
(for large ${\mit\Omega}$) 
is modified only quantitatively,
i.e. the value of the (two) critical fields increases.
The critical behaviour remains 
of the square-lattice Ising model universality class. 
If $\Delta{\mit\Omega}$ is equal to a half of the exchange Ising interaction, 
an additional quantum phase transition point appears at ${\mit\Omega}=0$. 
We have found a strong evidence 
that this critical point is characterised by the new set of exponents.
The other (two) critical points which 
correspond to the transition between the Ising and paramagnetic phases
are of the square-lattice Ising model universality class.
For larger $\Delta{\mit\Omega}$ 
the chain exhibits
four quantum phase transition points 
between the Ising and paramagnetic phases 
which belong to the square-lattice Ising model universality class.
We should stress that 
a part of the exponents have been obtained numerically.
Although, in principle,
the spin correlation functions required for the estimation of these exponents
can be calculated analytically
as in the uniform case, 
to our best knowledge, 
it has not yet been done. 
We have left this problem for future studies. 

\bigskip
{\small 
The authors are thankful to Johannes Richter and Oles' Zaburannyi 
for collaboration in the earlier stage of this study.
O. D. has enjoyed the hospitality of the guest program 
of the Max-Planck-Institut f\"{u}r Physik komplexer Systeme, Dresden
in the spring of 2004.}

\end{document}